\documentclass[12pt,a4paper,openany]{article}

\input{psfig}
\usepackage{graphics}
\usepackage{epsfig}

\begin{document}

\title{Calcium and synaptic dynamics underlying reverberatory activity in neuronal networks}
\author{
\centerline{Vladislav Volman$^{1,2,5,\dag}$, Richard C.
Gerkin$^{3,4}$, Pak-Ming Lau$^{3}$,}
\\
\centerline{Eshel Ben-Jacob$^{1,2}$, and Guo-Qiang Bi$^{3,4}$}
\newline
\\
\centerline{1. School of Physics and Astronomy, Raymond \& Beverly
Sackler Faculty of Exact Sciences,}
\\
\centerline{Tel-Aviv Univ., Tel-Aviv 69978, Israel}
\\
\centerline{2. Center for Theoretical Biological Physics,
University of California at San Diego,}
\\
\centerline{La Jolla, CA 92093, USA}
\\
\centerline{3. Department of Neurobiology, University of Pittsburgh School of Medicine,}
\\
\centerline{Pittsburgh, PA 15261, USA}
\\
\centerline{4. Center for Neuroscience and Center for the Neural Basis of Cognition,}
\\
\centerline{University of Pittsburgh, Pittsburgh, PA 15213, USA}
\\
\centerline{5. Computational Neurobiology Laboratory, The Salk
Institute for Biological Studies,}
\\
\centerline{La Jolla, CA 92037, USA}
\\
$^{\dag}$ communicating author (e-mail: volman@salk.edu)}

\maketitle
\baselineskip 18pt
\bibliographystyle{unsrt}

\section*{Abstract}
Persistent activity is postulated to drive
neural network plasticity and learning. To investigate its
underlying cellular mechanisms, we developed a biophysically
tractable model that explains the emergence, sustenance, and
eventual termination of short-term persistent activity. Using the
model, we reproduced the features of reverberating activity that
were observed in small (50-100 cells) networks of cultured
hippocampal neurons, such as the appearance of polysynaptic
current clusters, the typical inter-cluster intervals, the typical
duration of reverberation, and the response to changes in
extra-cellular ionic composition. The model relies on action
potential-triggered residual presynaptic calcium, which we suggest
plays an important role in sustaining reverberations. We show that
reverberatory activity is maintained by enhanced asynchronous
transmitter release from pre-synaptic terminals, which in itself
depends on the dynamics of residual presynaptic calcium. Hence,
asynchronous release, rather than being a "synaptic noise", can
play an important role in network dynamics. Additionally, we found
that a fast timescale synaptic depression is responsible for
oscillatory network activation during reverberations, whereas the
onset of a slow timescale depression leads to the termination of
reverberation. The simplicity of our model enabled a number of
predictions that were confirmed by additional analyses of
experimental manipulations.

\newpage
\section{Introduction}
Persistent neuronal activity likely underlies the operation of
working memory and other cellular and network functions
\cite{Durstewitz00,Wang01,Marder01}. The idea that reverberatory
activity is maintained in neuronal circuits by virtue of recurrent
excitation was first proposed by Lorente de N\'o and Hebb
\cite{Lorente33,HebbBook}, and has been developed into attractor
models \cite{Durstewitz00,Wang01}. Biophysically realistic models
have focused on a form of reverberation during which neurons in
the network fire in an uncorrelated fashion \cite{Wang01}.
Although the cellular mechanisms underlying the emergence of such
persistent activity in cortical networks are still being
unravelled \cite{McCormick03}, the slow kinetics of the
N-Methyl-D-Aspartate receptor (NMDAR) current has been proposed to
maintain persistent activity \cite{Wang01}, while negative
feedback from slow, activity-dependent $K^{+}$ currents has been
proposed to terminate it \cite{Compte03}. At the same time,
modelling studies of spontaneous activity in developing spinal
cord underscore the importance of multiple timescale
activity-dependent depression in episodic network oscillations
\cite{Tabak00}.

Recently we discovered that in small networks of cultured
hippocampal neurons, brief stimulation of one neuron can result in
correlated, reverberatory activity persisting for seconds
\cite{Lau05}. Using voltage-clamp recordings, rhythmic repetition
of polysynaptic current events can be monitored from single
neurons in the network. Typically, each event lasts for
$\approx100 msec$ and repeats at 5-10 Hz \cite{Lau05} (see also
figure \ref{fig-fig2}A). Such reverberatory activity is intriguing
in light of the role of network oscillations in information
processing and storage \cite{Sejnowski06}. Mechanistically, such
rhythmic reverberation may differ from the uncorrelated persistent
activity studied previously \cite{Wang01,Hansel01}. In particular,
these reverberations are maintained primarily by
amino-3-hydyroxy-5-methyl-4-isoxazolepropionic acid receptors
(AMPARs) and are virtually abolished by pharmacological
manipulations that block asynchronous transmitter release.
Asynchronous transmitter release is a fundamental property of
synaptic transmission, corresponding to an increased probability
of synaptic vesicle fusion that lasts for hundreds of milliseconds
following an action potential due to presynaptic residual calcium
elevation \cite{Lau05}. Since the rate of asynchronous release is
intimately linked to pre-synaptic calcium dynamics, these results
suggest that the latter might play an important role in sustaining
reverberations. Importantly, in most experiments no spontaneous
activity was observed, indicating that network reverberation in
small cultures is likely to be mediated by synaptic, rather than
intrinsic, mechanisms.

In this theoretical work we propose a minimal biophysical model
addressing the persistence and termination of rhythmic
reverberatory activity as observed in small neuronal circuits. The
model describes essential biophysical processes such as calcium
and synaptic dynamics that may underlie reverberation, yet is
controlled by a small parameter set, leading to testable
predictions. We demonstrate that the emergence of rhythmic
reverberatory activity can be explained by considering the
dynamics of residual presynaptic calcium. Specifically,
activity-dependent elevations of residual calcium lead to enhanced
levels of asynchronous transmitter release, thus enabling a
network to sustain reverberatory activity. We also show that
reverberations can be terminated by a slow timescale synaptic
depression. Furthermore, we associate model parameters with known
biophysical processes. Thus, we can compare model behavior as
these parameters change against experimental recordings obtained
under corresponding changes in physiological conditions. We
conclude that multiple timescale synaptic depression and the
dynamics of residual presynaptic calcium are essential mechanisms
underlying rhythmic reverberatory activity in small neuronal
networks.

\section{Methods}

\subsection{Cell culture and electrophysiology}
Island cultures of dissociated rat hippocampal neurons were
prepared as described previously \cite{Lau05}. Briefly, glass
cover-slips were coated with patterns of poly-l-lysine spots
($\approx1mm$ diameter) with custom-made rubber stamps.
Dissociated cells from embryonic day 17-18 rats were plated on the
cover-slips in 35-mm Petri dishes at densities of
$45\cdot10^{3}-90\cdot10^{3}$ cells per dish.  The culture medium
was DMEM (BioWhittaker) supplemented with $10\%$ heat-inactivated
bovine calf serum (HyClone), $10\%$ Hames F-12 with glutamine, 50
units/ml penicillin/streptomycin (Sigma) and 1x B-27 supplement
(Invitrogen). Cultures were used at 14-21 days in vitro when
reverberation was commonly observed in isolated networks of
$\approx50$ neurons on an island of monolayer glial cells.

Perforated whole-cell patch clamp recordings were carried out with
patch clamp amplifiers (Axopatch 700A, Axon Instruments) at room
temperature.  The pipette solution contained 136.5 mM
$K$-gluconate, 17.5 mM $KCl$, 9 mM $NaCl$, 1 mM $MgCl_{2}$, 10mM
Hepes, 0.2 mM EGT, and 200 μg/ml amphotericin B (pH7.3). The
external bath solution was a Hepes-buffered saline containing
(unless otherwise indicated) 150 mM $NaCl$, 3 mM $KCl$, 3 mM
$CaCl_{2}$, 2 mM $MgCl_{2}$, 10 mM Hepes, and 5 mM glucose (pH
7.3). Throughout the recording, the culture was perfused with
fresh bath solution at a constant rate of $\approx 1 ml/min$.
Synaptic transmission in networks cultured under these
experimental conditions is primarily characterized by a moderate
depression (PML and GQB, unpublished observations). Polysynaptic
current traces were recorded under a voltage clamp at a holding
potential of -70 mV. Voltage traces were recorded under a current
clamp. Stimulation pulses were 1-msec, 100-mV step depolarization
(voltage clamp) or 1-ms, 2-nA current injection (current clamp)
delivered at inter-stimulus intervals of at least 30 seconds.
Signals were filtered at 2 kHz and acquired by using a 16-bit
digitizing board (DigiData 3200, Axon Instruments) and processed
with the pClamp software (Axon Instrument) and custom MATLAB
(Mathworks) and IGOR Pro (Wavemetrics) programs. Reverberation
duration was defined as the time period from the point when the
rising phase of the first polysynaptic current (PSC) cluster
crossed a threshold ($\approx$ half of the average PSC amplitude)
to the time point when the falling phase of the last PSC cluster
crossed the threshold with no additional threshold crossing in the
next $500 msec$.

\subsection{The Tsodyks-Uziel-Markram model of synaptic transmission}
We begin with a simple, tractable model used by Tsodyks, Uziel and
Markram to describe short-term synaptic depression (henceforth
referred to as the TUM model) \cite{Tsodyks00}. In the TUM model,
the synaptic resource is assumed to be trafficking through any one
of 3 possible states: recovered ($X$), active ($Y$), or inactive
($Z$). The equations governing the exchange of transmitter between
these states are \cite{Tsodyks00}:
\begin{equation}\label{eq:eq1}
\begin{array}{ll}
\frac{dX}{dt}=\frac{Z}{\tau_{r}}-uX\delta(t-t_{spike}) \\
\frac{dY}{dt}=-\frac{Y}{\tau_{d}}+uX\delta(t-t_{spike}) \\
\frac{dZ}{dt}=\frac{Y}{\tau_{d}}-\frac{Z}{\tau_{r}}
\end{array}
\end{equation}

Here, $X, Y,$ and $Z$ are the fractions of synaptic resources in
the recovered, active, and inactive states, respectively.  We
assume that the overall amount of synaptic resource is a conserved
quantity, hence $X+Y+Z=1$, as can be seen from equation
\ref{eq:eq1}. The time-series $t_{spike}$ denote the arrival times
of pre-synaptic spikes, $\tau_{d}$ is the characteristic decay
time of post-synaptic currents (PSCs), and $\tau_{r}$ is the time
constant of recovery from synaptic depression. Since typically
cultured networks in our studies were small, we assume that there
are no synaptic delays. The value of $\tau_{d}$ is chosen to
reflect AMPAR current decay kinetics, since this postsynaptic
receptor class (but not NMDARs) are critical for reverberations in
hippocampal cultures \cite{Lau05}. The variable $u$ describes the
effective use of synaptic resources by an action potential
arriving at the presynaptic terminal(s), and it is constant for
excitatory synapses.

We assume that post-synaptic current (PSC) arriving at neuronal
somata via synapses, described by equation \ref{eq:eq1}, depends
linearly on the fraction of available synaptic resources. Hence,
an equivalent of synaptic conductance seen by a neuron is
$g_{syn}(t)=AY(t)$, where $A$ sets the scale for a density of
post-synaptic effectors such as glutamate receptors.

The synaptic current entering the soma is modelled in a
voltage-dependent fashion, and has the form
$I_{syn}(t)=-g_{syn}(t)(V-V_{r})$, where $V(t)$ is the
(time-dependent) membrane voltage (exact dynamics specified
below), and $V_{r}$ is the reversal potential of the corresponding
post-synaptic receptor type (we set $V_{r}=0 mV$ for AMPA
receptors). Summing the pre-synaptic contributions, the total PSC
entering the post-synaptic membrane of the $i$-th neuron is:
\begin{equation}\label{eq:eq2}
I^{i}_{syn}(t)=-\Sigma_{j}A_{ij}Y_{ij}(t)(V^{i}(t)-V^{ij}_{r})
\end{equation}

\subsection{Multiple time-scale depression}
\begin{figure}
\centerline{\resizebox{0.4\textwidth}{!}{\includegraphics{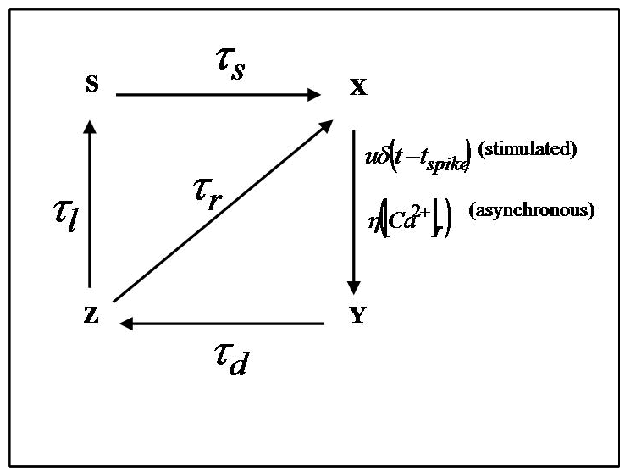}}}
\caption[]{\footnotesize Schematic representation of the model for
synaptic transmission used in this work. In the model,
neurotransmitter is trafficked through 4 functionally distinct
states. In the absence of stimulated release, some of the
transmitter is spontaneously transferred from the recovered state
($X$) to the active state ($Y$), with the rate of transfer
determined by the level of pre-synaptic residual calcium. After
the rapid ($\tau_{d}$) transition from active state to the
inactive state ($Z$), most of the transmitter follows the direct
recovery route (from $Z$ to $X$ within a time-scale of
$\tau_{r}$), whilst some fraction leaks to the super-inactive
state ($S$) within the characteristic time $\tau_{l}$. From there
it recovers to the $X$-state on a much slower time-scale
$\tau_{s}$. In addition to this basic route, action potentials
arriving at the synapse evoke stimulated release, during which the
utilized fraction of synaptic resources is determined by the
parameter $u$.}\label{fig-fig1}
\end{figure}

Synaptic dynamics are usually characterized by a diversity of
timescales, endowing synapses with short-term plasticity in
response to ongoing activity \cite{Hunter01,Gandhi03}. It has also
been demonstrated that hippocampal synapses utilize at least 3
distinct modes for vesicle recycling \cite{Gandhi03}, each marked
by a different timescale. To take into account the notion of
multiple time-scales for short-term synaptic plasticity on the
phenomenological level, we extended the TUM model by introducing
an additional synaptic state (called the super-inactive state).
Once in the inactive state (described by $Z$-variable), most of
the synaptic resource is directly transferred to the recovered
state ($X$-variable), while a small fraction leaks to the
super-inactive $S$-state, and then slowly reverts to the recovered
state. During periods of elevated neuronal activity, the
super-inactive state acts as a sink for synaptic resources thus
providing negative feedback at a slow timescale. The trafficking
of synaptic resource is shown schematically in figure
\ref{fig-fig1}. The augmented TUM model is thus:
\begin{equation}\label{eq:eq3}
\begin{array}{ll}
\frac{dX}{dt}=\frac{S}{\tau_{s}}+\frac{Z}{\tau_{r}}-uX\delta(t-t_{spike}) \\
\frac{dY}{dt}=-\frac{Y}{\tau_{d}}+uX\delta(t-t_{spike}) \\
\frac{dZ}{dt}=\frac{Y}{\tau_{d}}-\frac{Z}{\tau_{r}}-\frac{Z}{\tau_{l}} \\
\frac{dS}{dt}=\frac{Z}{\tau_{l}}-\frac{S}{\tau_{s}}
\end{array}
\end{equation}
The extent to which a super-inactive state influences persistent
activity is determined by the two parameters, $\tau_{l}$ and
$\tau_{s}$, which define the characteristic time of transmitter
deposition into the state and a typical time of recovery from the
slow depression, respectively. As shown below, varying these
parameters directly affects the duration of persistent activity in
the model network.

\subsection{The dynamics of residual $Ca^{2+}$}
The basic behavior of a synapse is well captured by the above
phenomenological model. However, more detailed synaptic
biophysical processes often turn out to be of great importance due
to their special role in the activity-dependent modulation of
synaptic transmission. Upon the arrival of an action potential,
the pre-synaptic terminal is depolarized, enabling calcium influx
through voltage-gated calcium channels. Pre-synaptic calcium
influx triggers a variety of processes, such as vesicular release
of neurotransmitter. As prolonged elevation of calcium
concentration can be detrimental to cell function, there exists
sophisticated machinery for calcium clearance. Most pre-synaptic
free calcium is rapidly extruded through the plasma membrane to
extra-cellular space \cite{Carafoli04,Zucker96}, but a small
portion is taken up, at a much slower rate, by pre-synaptic
organelles. The latter constitute pre-synaptic calcium stores, and
complex dynamics associated with these stores have been suggested
to play an important role in the regulation of synaptic function
\cite{FrielLNP}. Nonetheless, after each action potential a small
amount of free (not buffered) calcium is accumulated in the
cytosol of the synapse. This "residual calcium" has been
postulated to affect many of the characteristics of synaptic
transmission, most notably short-term synaptic dynamics and
asynchronous transmitter release \cite{Zucker96,Zucker02}.

Let us denote by $[Ca^{2+}]_{r}$ the concentration of residual
cytosolic calcium. We also assume that the extra-synaptic
concentration of calcium (formally denoted $[Ca^{2+}]_{o}$) is
constant. The values of $[Ca^{2+}]_{o}$ are on the order of $mM$,
while a typical concentration of residual calcium is in
sub-micro-molar region \cite{Zucker96}. The evolution of residual
pre-synaptic calcium is described by:
\begin{equation}\label{eq:eq4}
\frac{d[Ca^{2+}]_{r}}{dt}=\frac{-\beta([Ca^{2+}]_{r})^{n}}{k_{r}^{n}+([Ca^{2+}]_{r})^{n}}+I_{p}+\gamma log(\frac{[Ca^{2+}]_{o}}{[Ca^{2+}]_{r}})\delta(t-t_{spike})
\end{equation}

The first term in the above equation describes the action of
calcium pumps, extruding calcium from the cytoplasm back to the
extra-cellular space. The second term, $I_{p}$, corresponds to
passive flux of calcium into the cytosol. Finally, the third term
represents the influx of calcium into the cytosol upon arrival of
an action potential. This activity-dependent increase of residual
calcium occurs at each action potential arrival time $t_{spike}$,
and is driven by the electro-chemical gradient across the
membrane. The parameter $\gamma$ was tuned so as to result in
$\approx 100 nM$ per spike increment of residual calcium
\cite{Zucker96}. The steady-state pre-synaptic concentration of
residual calcium is therefore determined by setting the third term
equal to zero (no spikes), and setting
$\frac{d[Ca^{2+}]_{r}}{dt}=0$. This gives
$[Ca^{2+}]_{ss}=k_{r}\sqrt{\frac{I_{p}}{\beta-I_{p}}}$. Note that
because the dynamics of calcium in the synchronous phase of
transmission are much faster, the above model essentially
describes only the dynamics of residual calcium.

The exponent $n$ defines the degree of the cooperation required to
activate the pump, whereas the value of $k_{r}$ sets the
transition from decay kinetics in the low-concentration regime to
those in the high-concentration regime. For calcium extrusion
pumps, the values of these parameters have been estimated as
$k_{r}=0.4 \mu M, n=2$ \cite{Carafoli04}. Note also that while the
maximal rate of calcium extrusion, $\beta$, is constant, the
effective rate of pumping depends upon pre-synaptic calcium
concentration, thus ensuring regulatory feedback.

\subsection{Residual $Ca^{2+}$ and asynchronous release}
Although the exact biophysical details are unknown, evidence
suggests that the relation between the rate of asynchronous
release and the amount of residual calcium obeys the following
Hill function \cite{Ravin97,Kirischuk03}:
\begin{equation}\label{eq:eq5}
\eta([Ca^{2+}]_{r})=\eta_{max}\frac{([Ca^{2+}]_{r})^{m}}{k_{a}^{m}+([Ca^{2+}]_{r})^{m}}
\end{equation}
In the above equation, $\eta([Ca^{2+}]_{r})$ is the probability of
asynchronous release of transmitter (in a fixed time step), where
$\eta_{max}$ is the maximal rate of asynchronous transmission. The
Hill exponent, $m$, measures the degree of cooperation required in
order to activate the transmission. Higher values of $m$ make the
corresponding Hill curve steeper, and sharpen the distinction
between the behavior associated with low and high calcium
concentrations. Experiments performed on single synapses
\cite{Ravin97,Kirischuk03} estimate the Hill exponent to be in the
range $3 < m < 4$. In the simulation, we set $m=4$. In passing, we
note that such a high value of the Hill exponent (resulting in a
steep threshold) seems to be an important component for network
reverberations. For smaller values of the Hill exponent $(m=2)$,
less residual synaptic calcium is required to produce significant
levels of asynchronous activity. This can result in dynamical
instability of network dynamics, leading to spontaneous outbreaks
of reverberatory activity and to the de-coordination of population
dynamics (data not shown).

The threshold values of residual calcium needed to activate the
asynchronous release of transmitter vary from $<0.2 \mu M$ for
chromaffin cells \cite{Augustine92} to $>20 \mu M$ for retinal
bipolar nerve terminals \cite{Heidelberger94}.  This large
variability across different synapses might be attributed to the
existence of distinct $Ca^{2+}$ sensors or different molecular
organizations at the active zone \cite{Schneggenburger05}. Guided
by the reported importance of asynchronous transmitter release at
hippocampal synapses \cite{Hagler01,Lau05}, by the fact that
variations as small as $\Delta[Ca^{2+}]_{r}\sim 100 nM$ can
significantly increase the frequency of release, and by the
realization that the affinity for asynchronous release is
evidently higher than the affinity for $Ca^{2+}$ clearance, we
propose a sub-micro-molar value for $k_{a}$, and use $k_{a}=0.1\mu
M$ in the simulations.

\subsection{The enhanced (noisy) version of TUM model}
With the above relations between the rate of asynchronous release and concentration of pre-synaptic free calcium, the equations of synaptic transmission are modified, and take the following form:
\begin{equation}\label{eq:eq6}
\begin{array}{ll}
\frac{dX}{dt}=\frac{S}{\tau_{s}}+\frac{Z}{\tau_{r}}-uX\delta(t-t_{spike})-X\xi \\
\frac{dY}{dt}=-\frac{Y}{\tau_{d}}+uX\delta(t-t_{spike})+X\xi \\
\frac{dZ}{dt}=\frac{Y}{\tau_{d}}-\frac{Z}{\tau_{r}}-\frac{Z}{\tau_{l}} \\
\frac{dS}{dt}=\frac{Z}{\tau_{l}}-\frac{S}{\tau_{s}}
\end{array}
\end{equation}
where $\xi$ is an amplitude of the miniature synaptic event,
modelled as a random variable from a Gaussian distribution with
positive mean $\langle\xi\rangle$. We assume very weak variations
in the amplitude of miniature events, since these are believed to
be a release of a single vesicle, but still there could be some
variability in vesicle size. Such spontaneous events of stochastic
amplitude are generated with the calcium-dependent rate
$\eta([Ca^{2+}]_{r})$. The stochastic term is added in such a way
so as to satisfy the resource conservation constraint,
$X+Y+Z+S=1$. Note also that the quantity of spontaneously released
resource is proportional to the quantity of recovered synaptic
resource, $X$. Thus, the greater the recovered resource there is,
the greater the effective amplitude of asynchronous release.

\subsection{Neuronal Model}
Choosing a plausible description for neuronal dynamics constitutes
an important step in the construction of model that aims to
describe dynamical behavior of neuronal network. Single neuron
dynamics are usually studied using detailed compartmental models
\cite{Segev89}. This approach, however, becomes computationally
expensive when large assemblies of coupled neurons are considered.
A rule of thumb is to pick a simplified description of individual
neuron dynamics when large assemblies of cells are simulated, or
when a detailed membrane model is believed to be not critical to
the phenomena under study. As applied to the study of
reverberation in the cultured networks described here, we note
that: 1) since reverberating activity is marked by depolarization
of neuronal membrane and action potentials are of graded
magnitude, the dynamics and fluctuations of membrane voltage are
likely to be important; 2) studies of network dynamics and
synaptic kinetics become computationally expensive, especially
when longer periods of simulation (as is needed for the
exploration of parametric space) are needed. Guided by the desire
to retain a plausible model of neuronal dynamics, and at the same
time to have at hand an efficient (from a computational point of
view) system, we have chosen to describe the neuronal dynamics
with a simplified two-component, single compartment Morris-Lecar
model \cite{MorrisLecar81,RinzelErmentrout}:
\begin{equation}\label{eq:eq7}
\begin{array}{ll}
C\frac{dV}{dt}=-I_{ion}(V,W)+I_{ext}(t)\\
\frac{dW}{dt}=\theta\frac{W_{\infty}(V)-W(V)}{\tau_{W}(V)}\\
\tau_{W}(V)=(cosh(\frac{V-V_{3}}{2V_{4}}))^{-1}
\end{array}
\end{equation}
with $I_{ion}(V,W)$ representing the contribution of the internal
ionic $Ca^{2+}$, $K^{+}$ and leakage currents with their
corresponding channel conductivities $g_{Ca}$, $g_{K}$, and
$g_{L}$ being constant:
\begin{equation}\label{eq:eq8}
I_{ion}(V,W)=g_{Ca}m_{\infty}(V-V_{Ca})+g_{K}W(V-V_{K})+g_{L}(V-V_{L})
\end{equation}
The current $I_{ext}(t)$ represents all the external current
sources stimulating the neuron, such as signals received through
its synapses, glia-derived currents, artificial stimulation as
well as any noise sources.
\\
The equations for steady-state fractions of open potassium and calcium channels are:
\begin{equation}\label{eq:eq9}
\begin{array}{ll}
W_{\infty}(V)=\frac{1}{2}(1+tanh(\frac{V-V_{3}}{V_{4}}))\\
m_{\infty}(t)=\frac{1}{2}(1+tanh(\frac{V-V_{1}}{V_{2}}))\\
\end{array}
\end{equation}

The advantage of Morris-Lecar model is that it can be easily set
to describe either Type-I (saddle-node on invariant circle) or
Type-II (Hopf) dynamics of neuronal somata; the two types describe
the two qualitatively different ways in which the transition from
the quiescent to the spiking phase is accomplished
\cite{RinzelErmentrout}. Because experiments indicate that during
the reverberation phase neuronal somata are depolarized and
neuronal spike responses are of graded amplitude \cite{Lau05}, we
modelled the dynamics of individual neurons as Type-II. While
transmitter release in hippocampal neurons is coupled to sodium
spikes, rather than calcium spikes, the exact form of the neuronal
model is unimportant, provided that simulations of membrane
potential evolution roughly match experimental data. We were able
to obtain reverberatory activity in model networks using either
Type-II or Type-I neurons, indicating that the detailed form of
transition from  quiescent to spiking states is not crucial to
reverberation; nonetheless, we note that model reverberations with
Type-II neurons better resemble those observed in hippocampal
networks. Note also that although our model describes calcium
dynamics in both neuronal somata and synaptic boutons, the two
calcium concentrations (in soma and in synaptic bouton) are
different dynamical quantities. Thus only the synaptic calcium
variable is relevant for synaptic transmission, whereas the
somatic calcium variable is just for controlling membrane
potential dynamics as described by the Morris-Lecar model.

\begin{table}
\centering{
\begin{tabular}{|c|c|c|c|c|c|c|c|}
\hline
$g_{Ca}$ & $1.1\frac{mS}{cm^{2}}$ & $V_{L}$ & $-65 mV$ & $\theta$ & $0.2$ & $\langle\xi\rangle$ & $10^{-2}$ \\
\hline
$g_{K}$ & $2\frac{mS}{cm^{2}}$ & $V_{1}$ & $-1 mV$ & $\tau_{d}$ & $10 msec$ & $k_{r}$ & $0.4 \mu M$ \\
\hline
$g_{L}$ & $0.5\frac{mS}{cm^{2}}$ & $V_{2}$ & $15 mV$ & $\tau_{r}$ & $300 msec$ & $I_{p}$ & $0.11 \frac{\mu M}{sec}$ \\
\hline
$V_{Ca}$ & $100 mV$ & $V_{3}$ & $0 mV$ & $\tau_{s}$ & $10 sec$ & $\gamma$ & $80 \frac{\mu M}{sec}$ \\
\hline
$V_{K}$ & $-70 mV$ & $V_{4}$ & $30 mV$ & $k_{a}$ & $0.1 \mu M$ & $[Ca^{2+}]_{o}$ & $2 mM$ \\
\hline
$C$ & $1\frac{\mu F}{cm^{2}}$ & $\langle A\rangle$ & $3.41\frac{\mu A}{cm^{2}}$ & $n$ & $2$ & $m$ & $4$ \\
\hline
\end{tabular}
} \caption[]{\footnotesize Parameters used in simulations.}
\label{table-table1}
\end{table}

\subsection{Network connectivity and synaptic strength}
To investigate the characteristics of calcium-driven short-term
persistent activity in model networks, we simulated the dynamics
of a 60 neuron network. To comply with physiological data (unpublished observations), $10\%$
of the neurons were set to be inhibitory. Next, for each pair of
model neurons, we establish unidirectional synaptic connections
with probability $p_{0}$. After network topology was constructed, we drew
the absolute strength of synaptic connections between pairs of
neurons $(i,j)$ from a Gaussian distribution with positive mean.
Completely unbounded distributions of synaptic strength are not
physiologically plausible \cite{Bi01,vanRossum00}; in addition,
with very wide distributions, a model network is likely to exhibit
spontaneous outbreaks of reverberations (see Discussion). Since
our primary goal in this work is to explore the characteristics of
evoked reverberatory activity, we imposed boundaries on synaptic
strength, retaining only the values that satisfied the
localization constraint $0.8\langle A\rangle\le A\le 1.2\langle
A\rangle$. The synaptic connectivity, the neuronal threshold and
the averaged value of synaptic strength in the network were tuned
according to the following heuristic principle. Because of
geometrical constraints, neurons grown on isolated glial islands
can develop very strong synaptic connections consisting of large
numbers of boutons. In addition, experimental observations
\cite{Lau05} indicate that a single neuron fires 0-2 spikes during
the PSC cluster. Hence, we tuned the threshold and the averaged
synaptic strength so as to produce one post-synaptic spike upon
activation. The values of synaptic currents obtained in such a way
are compatible with the estimates for hippocampal synapses
\cite{Chen01}.

Finally, because reverberations in cultured networks are robust
and of finite duration even under the complete blockade of
GABAergic synapses \cite{Lau05} we mimic this condition by
nullifying the strength of all inhibitory synapses in our network.
Note that even under the complete blockade of inhibition, cultured
neurons in small networks fire at a moderate rate of $\sim 10
Hz$ during reverberation. This is much lower than the high
frequency firing characteristic of paroxysmal discharges during
epileptic seizure.

\section{Results and Discussion}

\subsection{The emergence of network reverberations}
\begin{figure}
\centerline{\resizebox{0.5\textwidth}{0.5\textheight}{\includegraphics{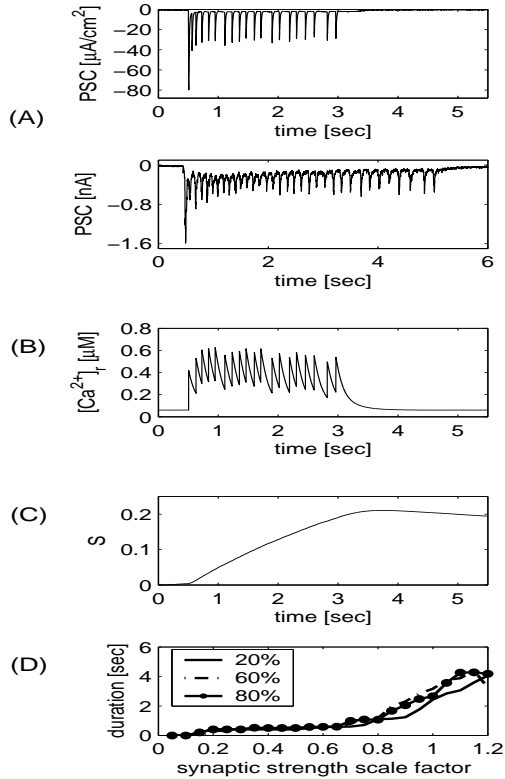}}}
\caption[]{\footnotesize Activity characteristics of a model
neuron. A) Upper panel: a postsynaptic current (PSC) trace of a
sample neuron from a model network demonstrates the emergence of
reverberating activity. The activity persists for several seconds,
and spontaneously terminates as a result of slow time-scale
depression. Lower panel: For comparison, a PSC trace recorded at a
voltage-clamped neuron in a cultured network is shown to
illustrate reverberatory network activity in response to a 1-ms
stimulation of another neuron. As previously reported
\cite{Lau05}, this reverberatory activity takes the form of
polysynaptic current clusters; each cluster lasts for $\sim 50
msec$, and the temporal gap between two adjacent clusters is $\sim
150 msec$. Note the overall similarity between the simulated and
experimentally recorded traces. B) Pre-synaptic residual calcium
at a sample model synaptic terminal. In the absence of stimulated
activity, the concentration of pre-synaptic calcium relaxes to
$[Ca^{2+}]_{ss}\sim 60 nM$. Action potentials arriving at the
synaptic terminal lead to the transient elevation of pre-synaptic
calcium levels, with the typical values reaching $0.8 \mu M$. This
elevated calcium is responsible for the enhanced level of
asynchronous release, which serves to sustain the network in the
reverberating state. C) During the reverberatory phase, there is a
constant deposition of synaptic resource into the super-inactive
depression state, from where it recovers with a very slow (several
seconds) time constant. Consequently, the fraction of synaptic
resource in the slow depression state eventually becomes high
enough to make synaptic transmission ineffective, leading to
termination of the network reverberation. D) Dependence of evoked
reverberatory activity on the characteristics of synaptic
strength. Mean strength of model synapses has been scaled relative
to the value used in the rest of the simulations. Networks with
low averaged synaptic strength responded with a single PSC
cluster. A transition from non-reverberatory to reverberatory
phase occurred at $0.65$ of the value used in simulations.
Variation of constraints on strength distribution (boundaries
located $20\%$, $60\%$ and $80\%$ away from mean value) led
qualitatively to the same profile of phase transition. Simulations
in (A)-(C) were with $\langle A\rangle=3.41\frac{\mu A}{cm^{2}}$,
and data points were averaged over 10 statistically independent
realizations.} \label{fig-fig2}
\end{figure}

We begin with general observations regarding the dynamical
behavior of the model network. We found that in a typical
simulation (parameter choices specified in Table
\ref{table-table1}) the network response to brief stimulation (5
msec duration) of one neuron is characterized by the appearance of
a recurrent, excitatory reverberation. This is seen clearly from
figure \ref{fig-fig2}A (top), in which the total post-synaptic
current experienced by a sample neuron (in voltage clamp) from the
model is shown. The temporal profile of the post-synaptic current
resembles that observed in small networks of cultured neurons
(figure \ref{fig-fig2}A, lower panel). Similar to the experimental
results, the reverberating activity in the model network
terminates after several seconds. We subsequently show that both
the temporal gaps between the adjacent current clusters and the
duration of network reverberation are controlled by specific
biophysical parameters.

In figure \ref{fig-fig2}B, we show the dynamics of pre-synaptic
residual calcium in the model. Each action potential invading the
pre-synaptic terminal during a PSC cluster contributes to the
build-up of such calcium. When a PSC cluster is terminated (due to
short-term synaptic depression of evoked transmitter release),
residual calcium levels are sufficiently elevated to generate
abundant asynchronous transmitter release, consistent with
observations in various types of synapses \cite{Goda94,Atluri98}.
While the concentration of presynaptic calcium slowly decays due
to pump-related active extrusion, the fraction of synaptic
resource recovered from depression simultaneously rises. At a
critical point, synaptic resource recovery permits asynchronous
post-synaptic current to trigger the next PSC cluster.

\begin{figure}
\centerline{\resizebox{0.4\textwidth}{!}{\includegraphics{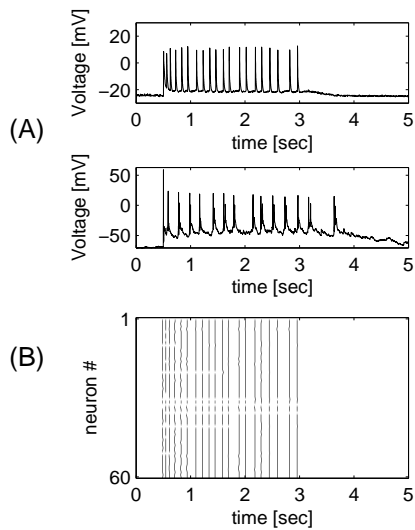}}}
\caption[]{\footnotesize Membrane voltage fluctuations in a
reverberatory model network. A) Upper panel: Membrane voltage
trace of a sample neuron in the model network reveals that the
neuron fires 1 spike during each PSC cluster. Between the
clusters, the neuronal membrane potential fluctuates due to the
enhanced levels of asynchronous release. Lower panel: Sample
voltage trace recorded in a current clamp from a neuron in a
cultured network \cite{Lau05}. Reverberatory network activity is
detected as membrane depolarization. During each PSC cluster, a
neuron typically fires 0-2 spikes. Also, note the sub-threshold
fluctuations of membrane potential between spikes. B) The
reverberatory behavior at the network level is reflected in a
raster plot of network activity (black marks correspond to
neuronal firing). The raster plot reveals that during the PSC
clusters, most of the population is firing in a highly correlated
manner. Hence, PSC clusters in the model network represent
temporally localized events. The temporal resolution (size of each
bin) is $10 msec$.}\label{fig-fig3}
\end{figure}

Reverberatory activity in the model network terminates due to the
accumulation of synaptic resource in the super-inactive state (as
shown in figure \ref{fig-fig2}C). Thus our model demonstrates that
presynaptic dynamics alone have the capacity to account for
important features such as the initiation, continuance, and
eventual termination of reverberatory activity similar to that
observed in networks of hippocampal neurons \cite{Lau05}. This
conclusion is further supported by the raster plot of network
activity in the model, shown in figure \ref{fig-fig3}B. The
appearance of PSC clusters reflects the existence of short time
windows during which the spiking activity of the network
population is highly correlated. Note that there are no spikes in
the quiescent periods between clusters, indicating that
asynchronous release alone might be sufficient to provide the
driving force for the initiation of the next cluster. Somatic
calcium imaging data also indicates that the inter-PSC cluster
periods contain few or no spikes (PML and GQB, unpublished
observations).

Note that the 4-state model of synaptic resources utilized by us
is crucial for persistent activity. In our model, a network has to
accumulate a certain "critical mass" of available
neuro-transmitter in order to sustain its activity. Using a
3-state model (with one depression state) and a very long recovery
time-scale will lead either to a failure to evoke reverberations
(for strong depression), or to a transition to an asynchronous
mode of activity (for weak depression). Neither corresponds to
experimentally observed reverberations of finite duration.

To test how the emergence of reverberatory activity in model
networks is determined by the overall synaptic strength
distribution in the network, we evaluated the duration of evoked
reverberations for a range of systematically varied values of the
mean strength of model synapses. Figure \ref{fig-fig2}D
demonstrates that, for low values of mean synaptic strength,
stimulation failed to induce reverberations in a model network.
For higher values of synaptic strength, short reverberations (1-2
PSC clusters) were obtained. Finally, a transition to
reverberations lasting for several seconds occurred slightly below
the point that we used in the rest of the model simulations.
Increasing the variance of the synaptic strengths (replacing the
$20\%$ constraint with the $80\%$ constraint, i.e using
$0.2\langle A\rangle\le A\le 1.8\langle A\rangle$) did not lead to
any qualitative changes in the profile of transition from
non-reverberatory to reverberatory phases (figure
\ref{fig-fig2}D). These results indicate that in our model
reverberation is a robust phenomenon with respect to variations in
synaptic strength distributions. Meanwhile, the mean connection
strength in the network plays a critical role in the emergence of
reverberation, consistent with experimental observations
\cite{Lau05}.

The membrane voltage profile for a sample model neuron (figure
\ref{fig-fig3}A, upper panel) shows that the phase of persistent
network activity is characterized by neuronal membrane
depolarization (driven by polysynaptic inputs). This observation
is again in qualitative agreement with experimental findings
(figure \ref{fig-fig3}A, lower panel), yet quantitatively the
range of the model's membrane potential values appears to be off
by a constant factor from those of cultured neurons. This
discrepancy follows from model neuron selection (ML), and one
might argue that selecting another model would lead to
qualitatively different results. To test whether this is the case,
we performed model simulations with both type-I and type-II
neurons (selecting two versions of Morris-Lecar model). For both
choices, we observed network reverberation (data not shown).
Hence, the choice of a particular spiking neuron model does not
appear to be critical for the emergence of reverberations.

\subsection{Reverberation duration in small networks is determined by the rate of slow
depression and the rate of pre-synaptic calcium extrusion} Earlier
studies have explored the mechanisms responsible for the existence
of various types of persistent activity in large neuronal networks
across different preparations. For example, spindle oscillations
in a large-scale thalamic model were found to emerge from
reciprocal interactions between excitatory and inhibitory
populations \cite{Golomb94}. On the other hand, the synaptic basis
of cortical persistent activity is thought to be activation of
NMDA receptors \cite{Wang01}. However, inhibition and NMDA-like
excitation were shown to be not necessary for network
reverberations in small hippocampal cultures \cite{Lau05}.
Instead, the ability of a model network to exhibit reverberatory
activity depends on the factors that determine the readiness of
the network to generate a new PSC cluster at each time-point.
Examples of such time-dependent factors are effective synaptic
strength and distance to action-potential threshold. The
parameters that determine whether sequential PSC cluster will
continue to be generated, and that control reverberation duration,
are likely to be related to underlying biophysical processes. Two
prominent examples are the rate of transmitter leakage into
super-inactive state, and the maximal rate of pre-synaptic
residual calcium clearance. Since the exact physiological values
for these two rates are not known, we systematically varied them
to investigate their effects on the duration of reverberating
activity in the model network.

\begin{figure}
\centerline{\resizebox{0.4\textwidth}{!}{\includegraphics{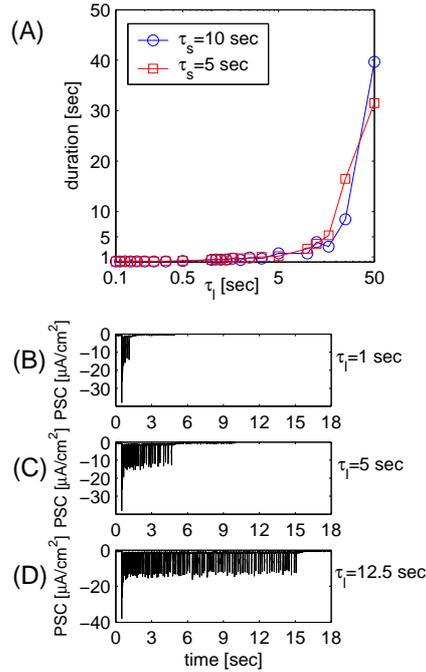}}}
\caption[]{\footnotesize Dependence of reverberation duration on
the rate of slow depression. A) Duration as a function of the time
constant of slow depression, $\tau_{l}$. As the value of
$\tau_{l}$ is decreased (for a given value of recovery time),
there is a faster leak rate of transmitter into the super-inactive
state, from which it recovers on a time-scale of several seconds.
Consequently, after a certain time the amount of resource in the
recovered state is insufficient to sustain the network in the
reverberating state. For high rates of slow depression the
deposition rate is so fast that the activity of a network
terminates after a few PSC clusters. The effect of varying the
recovery time is clearly assessed when comparing the results for
different values of $\tau_{s}$  (red curve for $\tau_{s}=5 sec$
and blue curve for $\tau_{s}=10 sec$). For certain values of
typical time, the networks with shorter recovery time can exhibit
longer reverberations. All simulations have been performed with
$\beta=5\cdot10^{-3}$, and all data points are averaged over 10
statistically independent realizations. B,C,D) Sample PSC traces
for $\tau_{l}=1 sec$, $\tau_{l}=5 sec$ and $\tau_{l}=12.5 sec$
respectively, are shown to illustrate the qualitative dependence
of reverberation duration on the rate of slow
depression.}\label{fig-fig4}
\end{figure}

As figure \ref{fig-fig4} demonstrates, lowering the rate of
synaptic resource leakage into the super-inactive state leads to
an increase in reverberation duration. When $\tau_{s}<\tau_{l}$,
the recovery of transmitter from slow synaptic depression is fast
enough to enable the network to sustain reverberation
indefinitely. At the other extreme, making $\tau_{l}$ very rapid
virtually abolishes reverberations.

\begin{figure}
\centerline{\resizebox{0.4\textwidth}{!}{\includegraphics{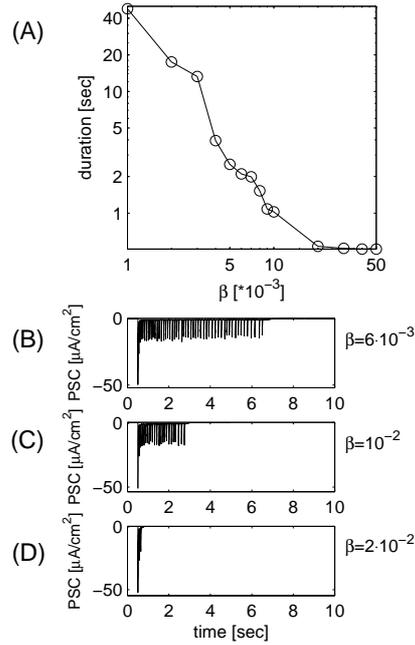}}}
\caption[]{\footnotesize Dependence of reverberation duration on
the maximal rate of pre-synaptic calcium clearance, $\beta$. A)
Duration as a function of the maximal rate of $Ca^{2+}$ clearance.
Higher values of calcium clearance are more effective at reducing
the concentration of residual presynaptic calcium. Consequently,
the levels of asynchronous release are lower, leading to earlier
termination of network reverberation. B,C,D) Sample PSC traces for
$\beta=6\cdot10^{-3}$, $\beta=10^{-2}$ and $\beta=2\cdot10^{-2}$
respectively, are shown to illustrate the effect of varying the
value of the calcium clearance rate. All simulations were
performed with $\tau_{l}=5 sec$, and data points were averaged
over 10 statistically independent realizations. Note that the
actual duration of reverberation shown in (B) is longer than the
corresponding median value (shown in (A)), indicating that
individual realizations exhibit fluctuations (due to slightly
different connectivity schemes).}\label{fig-fig5}
\end{figure}

We next proceeded to determine the effect of maximal calcium
extrusion rate on reverberation duration. As shown in figure
\ref{fig-fig5}, increasing the maximal rates of calcium clearance,
$\beta$, (i.e., faster effective extrusion time) decreases the
rate of asynchronous release, shortening reverberation duration.
However, the extent to which a calcium clearance rate can affect
the duration of persistent activity is limited, because the
maximal admissible duration is determined by the value of
$\tau_{l}$ (typical time of transmitter deposition into the slow
depression state).

\subsection{Synchronous vs. asynchronous release}
\begin{figure}
\centerline{\resizebox{0.4\textwidth}{!}{\includegraphics{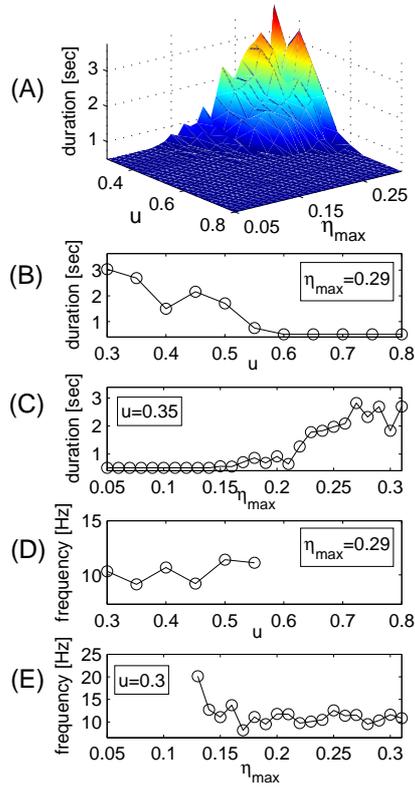}}}
\caption[]{\footnotesize Dependence of reverberation
characteristics on the interplay between synchronous and
asynchronous transmitter release. A) Duration of reverberation as
a function of maximal frequency of asynchronous release
$\eta_{max}$, and stimulated resource utilization parameter $u$. A
general trend is an increase in the duration of reverberating
activity as the value of $\eta_{max}$ is increased, and a decrease
in the duration of reverberation as the value of $u$ is increased.
This qualitative dependence on the two parameters is further
demonstrated in (B,C), where two cross-sections of the duration
surface (for constant $\eta_{max}$ and constant $u$) are shown.
D,E) The frequency of PSC cluster appearance depends only weakly
on the maximal frequency of asynchronous release and the resource
utilization parameter, consistent with experimental results (see
Results). The simulations were performed with $\tau_{l}=5 sec$,
$\beta=5\cdot10^{-3}$ and all data points were averaged over 10
statistically independent realizations.} \label{fig-fig6}
\end{figure}

In the model network, the sustainability of a reverberation
depends on the interplay between synchronous and asynchronous
release of neurotransmitter. Hence, it is of interest to study the
influence of model parameters describing these two kinds of
release on the emergence and characteristics of reverberations. To
this end, we performed model simulations for systematically varied
values of the resource utilization parameter, $u$, and the maximal
rate of asynchronous release, $\eta_{max}$ (figure
\ref{fig-fig6}A). The values of $\tau_{l}$ and $\beta$ were held
constant throughout these simulations, so as to fix the maximal
admissible reverberation duration.

The value of the resource utilization parameter contributes to the
magnitude of the postsynaptic event following the arrival of an
action potential at the pre-synaptic terminal. Consequently, the
resource utilization parameter is one of the factors determining
the ability of a network to evoke the next PSC cluster. Low
utilization of synaptic resource (small values of $u$) might
result in insufficient excitation to generate a PSC cluster. On
the other hand, high utilization can rapidly deplete the pool of
transmitter, thus leading to the termination of a reverberation
due to rapid accumulation of a synaptic resource in inactive
states. In figure \ref{fig-fig6}B we show the qualitative change
in duration of reverberation as the value of $u$ is varied.
Typically, reverberation duration is inversely related to the
value of $u$. However, below a lower critical value for $u$, the
synchronous phase of evoked release becomes insignificant
(relative to the asynchronous mode), and networks operate in an
uncoordinated regime, in which the network's activity is marked by
relatively high-frequency non-coordinated neuronal discharges
(data not shown).

We next evaluated the effect of varying $\eta_{max}$, the maximal
rate of asynchronous release, on the duration of reverberatory
activity. As shown in figure \ref{fig-fig6}C, increasing
$\eta_{max}$ typically increases the duration of a reverberation.
However, when $\eta_{max}$ is above a critical value, asynchronous
release dominates over synchronous release, and the network again
switches to an uncoordinated mode of activity (data not shown). At
the other extreme, for $\eta_{max}$ below a lower critical value
reverberations are virtually abolished, because the level of
asynchronous release is too low to provide neurons with sufficient
tonic synaptic drive for subsequent PSC clusters to be generated.

The specific balance between synchronous and asynchronous release
could also affect the duration of the temporal gaps between
adjacent PSC clusters.  To test whether this is the case, we
computed the average frequency of reverberation (defined as
$ICI^{-1}$, where an inter-cluster-interval ICI is computed as the
mean interval between the peaks of adjacent PSC clusters) for
different values of the resource utilization parameter $u$ and the
maximal frequency of asynchronous release $\eta_{max}$. Figures
\ref{fig-fig6}D,E show the trend in reverberation frequency as a
function of synchronous and asynchronous release parameters. While
frequency tends to increase with increasing $\eta_{max}$, the
dependence is much weaker than that of reverberation duration.

\subsection{Testing the model with strontium experiments}
The effects of asynchronous release and pre-synaptic calcium were
studied experimentally by replacing calcium with strontium in the
extra-cellular medium. Strontium is known to de-synchronize
transmitter release, thus enhancing the asynchronous mode of
release. For experiments in which strontium partially replaced
calcium, the amplitude of the current at the peak of each PSC
cluster (synchronous phase) was reduced, while the amplitude of
the current between peaks (asynchronous phase) was enhanced
\cite{Lau05}. Furthermore, reverberations lasted longer. These
results imply that by manipulating the "trade-off" between the
synchronous and asynchronous phases of synaptic transmission it is
possible to influence the characteristics of reverberatory
activity.

\begin{figure}
\centerline{\resizebox{0.4\textwidth}{!}{\includegraphics{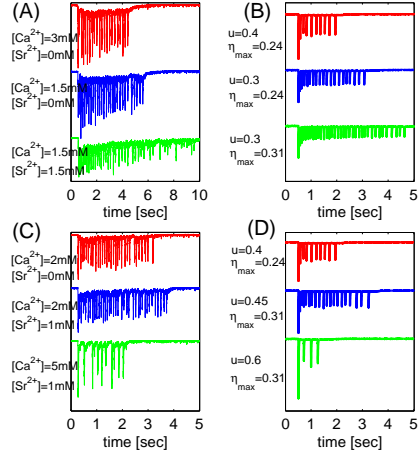}}}
\caption[]{\footnotesize Generic comparison between the behavior
of model and experimental networks in response to changes in ion
concentrations. A) Representative reverberation traces recorded
from the same network under different concentrations of $Ca^{2+}$
and $Sr^{2+}$ - $[Ca^{2+}]=3mM,[Sr^{2+}]=0mM$ (red),
$[Ca^{2+}]=1.5mM,[Sr^{2+}]=0mM$ (blue), and
$[Ca^{2+}]=1.5mM,[Sr^{2+}]=1.5mM$ (green). B) Model reverberations
obtained with the identification of release parameters as
described in the text: $u=0.4,\eta_{max}=0.24$ (red);
$u=0.3,\eta_{max}=0.24$ (blue); and $u=0.3,\eta_{max}=0.31$
(green). C) Experimental traces recorded from another network
under - $[Ca^{2+}]=2mM,[Sr^{2+}]=0mM$ (red);
$[Ca^{2+}]=2mM,[Sr^{2+}]=1mM$ (blue); and
$[Ca^{2+}]=5mM,[Sr^{2+}]=1mM$ (green). D) Model reverberations
obtained for - $u=0.4,\eta_{max}=0.24$ (red),
$u=0.45,\eta_{max}=0.31$ (blue), and $u=0.6,\eta_{max}=0.31$
(green). All model simulations were performed with $\tau_{l}=5
sec,\beta=5\cdot10^{-3}$. These data illustrate that experimental
alterations of the rate of both synchronous and asynchronous
release dictate the properties of reverberation, and that these
features are recapitulated by manipulations of the corresponding
parameters in the model.}\label{fig-fig7}
\end{figure}

In the framework of the model network, strontium-induced partial
suppression of stimulated synaptic transmission would correspond
to a smaller value of resource utilization parameter, $u$. On the
other hand, enhancement of asynchronous release corresponds to
increasing the value of $\eta_{max}$ (maximal release frequency).
In order to test the predictions of the model, we made the
simplifying assumption that the synchronous and asynchronous modes
of transmitter release can be manipulated independently. Thus,
given the approximate correspondence between model parameters and
biophysical processes, we compared the experimentally observed
reverberation traces under different conditions with their
analogues in the model. As is seen from figure \ref{fig-fig7},
these parameter changes in the model correctly reproduce the
experimental effects of strontium replacing calcium. Namely, upon
treatment with "strontium", the average amplitude of the model
network's PSC clusters decreases, while the duration of
reverberations is markedly increased. An increase in the
concentration of calcium has the opposite effect. In addition, the
model predicts that the average temporal gap between adjacent PSC
clusters should decrease moderately in the presence of strontium
(i.e. the frequency at which the network reverberates should grow
higher) (13 Hz in "strontium" vs. 10 Hz for "control").
Quantitative analysis of reverberation experiments indeed shows a
slight increase in reverberation frequency, consistent with this
prediction ($6.61\pm0.13$ Hz in strontium vs. $5.95\pm0.21$ Hz for
controls, $p<0.01$, Student's t-test).

\subsection{Spontaneous outbreaks of reverberation and uncoordinated population activity}
\begin{figure}
\centerline{\resizebox{0.4\textwidth}{!}{\includegraphics{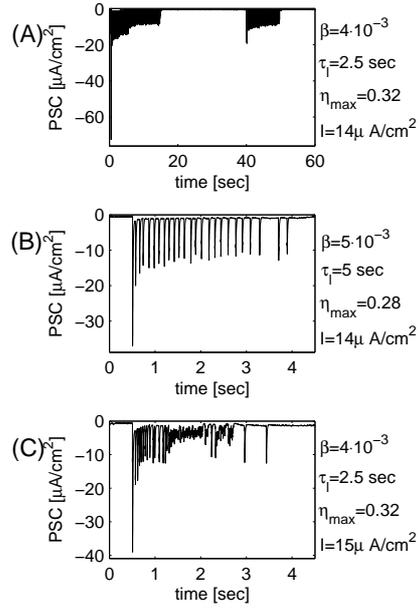}}}
\caption[]{\footnotesize The effect of enhanced excitability on
network activity. A) Spontaneous outbreaks of reverberatory
activity could be observed when the constraint on synaptic
strength was relaxed. Parameters used:
$u=0.4,\eta_{max}=0.28,\tau_{l}=5 sec,\beta=5\cdot10^{-3}$ and
$I=14\frac{\mu A}{cm^{2}}$.  Mean synaptic strength is as
indicated in the Table, but no constraint was used to obtain the
values of absolute synaptic strength. B) In a typical simulation,
a brief stimulation led to reverberatory activity lasting for
several seconds. Parameters used:
$u=0.4,\eta_{max}=0.32,\beta=4\cdot10^{-3},\tau_{l}=2.5
sec,I=14\frac{\mu A}{cm^{2}}$. Synaptic strength distribution was
constrained as described in Methods. C) With an elevated value of
background current ($I=15\frac{\mu A}{cm^{2}}$, other parameters
are the same as in (B)), simulation with an otherwise identical
set of model parameters resulted in the appearance of periods of
uncoordinated activity.}\label{fig-fig8}
\end{figure}

Enhanced network excitability raised the chances of observing
spontaneous network reverberation. When we relaxed the constraints
on the variance of synaptic strength, the model network could
exhibit spontaneous reverberations (in the absence of any
stimulus, figure \ref{fig-fig8}A). This observation can be
attributed to the existence of a few very strong synaptic
connections that are themselves capable of initiating the first
PSC cluster of a reverberation with the help of a modest
background current. Indeed, manipulations that increase synaptic
strength in cultured networks can also lead to spontaneous
reverberation in cultured networks (RCG and GQB, unpublished
data). In model networks we could control the appearance of
reverberations by varying the relevant parameters. Normally
growing networks in culture may utilize biophysical pathways to
regulate their excitability in an activity-dependent manner, thus
ensuring that reverberations have the appropriate balance of
rarity, PSC cluster frequency, and duration. Examples of such
mechanisms could be spike timing dependent plasticity \cite{Bi01},
activity-dependent scaling of synaptic weights
\cite{Turrigiano98}, or modulation from adjacent astro-glia
\cite{Angulo04}. Further research should delineate the
contribution of these to the ability of a network to sustain
healthy reverberations.

Although typically the reverberations in our model network
followed the patterns of recorded activity quite well, we observed
some potentially interesting distinctions between simulated and
real networks. Namely, in some cases the simulated reverberations
were interrupted by periods of relatively dense synaptic activity.
During these windows, the electrical activity of the neuronal
population was uncoordinated. Such uncoordinated behavior was not
observed in cultured networks, and therefore has drawn our
attention. Upon closer examination, we found that uncoordinated
activity could be induced in networks with relatively elevated
neuronal excitability. In addition to changes in the variance of
the synaptic distribution described above, the excitability can
also be affected by a number of other parameters - for example,
the value of the background current, the maximal frequency of
asynchronous transmitter release, etc. As an example, elevating
the value of constant background current to $I=15\frac{\mu
A}{cm^{2}}$ (instead of $I=14\frac{\mu A}{cm^{2}}$, as used in
other simulations and as specified in Table \ref{table-table1})
led to the appearance of uncoordinated activity under otherwise
"normal" conditions, as shown in figures \ref{fig-fig8}B,C.

To test the idea that additional biophysical mechanisms might
down-regulate these otherwise unstable network states, we modified
model synapses to include an additional, high capacity low
affinity calcium extrusion pump \cite{Carafoli04,FrielLNP}. Equation \ref{eq:eq4} now reads:
\begin{eqnarray}\label{eq:eq12}
\frac{d[Ca^{2+}]_{r}}{dt}=\frac{-\beta([Ca^{2+}]_{r})^{n}}{k_{r}^{n}+([Ca^{2+}]_{r})^{n}}-\frac{0.05([Ca^{2+}]_{r})^{2}}{4+([Ca^{2+}]_{r})^{2}}+I_{p}+\nonumber\\
\space+\gamma
log(\frac{[Ca^{2+}]_{o}}{[Ca^{2+}]_{r}})\delta(t-t_{spike})
\end{eqnarray}
Since the affinity of an additional pump is much lower than
typical levels of synaptic residual calcium, it is expected to
have little effect during "clean" reverberations. However, as seen
from figure \ref{fig-fig9}, the residual calcium in model synapses
is rather high during outbreaks of uncoordinated activity. During
these outbreaks, the number of neurons active during a narrow time
window (roughly corresponding to the temporal "localization" of
PSC cluster) falls low. In this situation, the addition of a low
affinity pump proves beneficial, as it can help to keep the
network in the "normal" reverberating state (figure
\ref{fig-fig9}).

\section{Conclusion and Outlook}
In this work, we provide a simple, yet biophysically tractable
modelling framework that aims to explain the emergence,
persistence, and eventual termination of reverberatory activity
observed in small networks of cultured neurons. Using this
framework, we reproduced the salient features observed in the
activity of small cultured networks - network reverberations
manifested themselves in the appearance of polysynaptic current
(PSC) clusters, with a typical cluster width and inter-cluster
separation matching those observed in experiments. The appearance
of clusters corresponded to the short periods of time during which
the activity of networks elements was highly coordinated.

This work also offers new insight into the mechanisms by which a
network could interdigitate highly coordinated population activity
with periods of silence. In order for activity to re-emerge after
the silent periods, asynchronous neurotransmitter release must be
sufficient to maintain the excitability of the network, while the
recovery of neurotransmitter from short-term depression eventually
triggers a new round of activity. This represents a biophysical
implementation of the hypothesis that rapidly decaying negative
feedback, combined with slowly decaying positive feedback, could
maintain persistent activity in neuronal networks \cite{Golomb06}.
This may represent a general mechanism by which such networks can
sustain reverberation.

Cultured networks are also known to exhibit bouts of activity
called synchronized bursting events (SBEs) \cite{Segev01}.
However, the reverberations seen here in small networks differ
from the SBEs observed in large networks in at least two aspects:
a) the width of a typical SBE is larger (by a factor of 2-4) than
the width of a typical PSC cluster, and b) the separation between
a pair of adjacent SBEs is several orders of magnitude larger than
the inter-PSC-cluster separation. This observation hints that the
mechanism underlying the generation of reverberatory phenomena in
small networks differs from the one responsible for the generation
of SBEs in larger networks. Exactly how these mechanisms and their
expression depend on network size is a question that should be
resolved by further experimental and modelling studies.

It should be noted that our description of network dynamics
assumes that a single action potential arriving at the synaptic
terminal can evoke a spike in some of its post-synaptic targets
(and generate a PSC cluster in an avalanche-like manner). Contrary
to this, experiments indicate that there is a great deal of
indeterminacy in neuronal responses; single stimuli were not
always successful in evoking reverberatory activity \cite{Lau05}.
Interestingly, enhanced activation of reverberation could be
obtained with paired-pulse stimulation, with each pulse separated
by 200-400 msec. Such paired-pulse stimuli produced greater levels
of asynchronous release than one pulse alone. This observation
might indicate that physiologically relevant input stimuli might
drive the synapse into a resonant state with respect to
presynaptic residual calcium dynamics, ultimately imparting the
network with the level of excitability needed to sustain a
reverberation.

In the present work, we have assumed that evoked synaptic
transmission is "faithful" (i.e. each incoming action potential
reliably leads to transmitter release), whereas asynchronous
transmitter release was modelled as a stochastic process. In
reality, transmitter release is inherently stochastic; yet, due to
the relatively large number of vesicles involved in evoked release
in these cultures, this process might be close to deterministic.
Were the variability of evoked transmitter release incorporated in
our model, it would only lead to a better match with experiments
(see previous paragraph). However, such a level of description
should include detailed modelling of presynaptic biophysics, and
might become computationally much more complex for network
simulations. Meanwhile, our hybrid system of deterministic evoked
synaptic transmission and stochastic asynchronous release should
be considered as a first approximation towards the more realistic
models of calcium-driven network reverberations.

An important hypothesis concerning persistent network activity
suggests that the slow kinetics of postsynaptic NMDA receptors is
necessary for such activity to be sustained in an asynchronous
fashion \cite{Wang01}. Our work, along with recent experiments
\cite{Lau05}, indicates an alternative mechanism whereby
asynchronous release of neurotransmitter from presynaptic
terminals, also with a slow kinetics, is a critical factor for the
emergence of rhythmic reverberations in small neuronal circuits.
It is important to note that our results do not contradict the
role of NMDARs - rather, given the sustained membrane
depolarization (which relieves $Mg^{2+}$ block of NMDARs) during
reverberation, it is plausible that NMDAR activation (a
postsynaptic process that is enhanced by membrane depolarization)
might cooperate with asynchronous release (a presynaptic process
that is enhanced by repeated activation) to increase the synaptic
drive onto neurons during reverberations. While experimental
results show that totally blocking NMDAR-mediated transmission can
weaken reverberation \cite{Lau05}, partially supporting this view,
a more systematic analysis could be of interest.

Just as NMDA receptors increase network excitability, GABAergic
inhibitory neurons can prevent reverberations or shorten their
duration by providing negative feedback. In the present model, we
have considered network reverberations in the absence of
inhibition in order to delineate the importance of presynaptic
mechanisms of recurrent excitation. In so doing, we relied on the
experimental observation that in small networks normal
reverberatory activity is present (and in fact enhanced) even
under the complete blockade of GABAergic synapses \cite{Lau05}.
Although beyond the scope of the current study, it is likely that
in larger networks with stronger excitatory connections,
inhibition plays critical roles in balancing network dynamics. In
such cases, presynaptic residual calcium may also be important in
modulating the dynamics of GABAergic transmission as well.

\begin{figure}
\centerline{\resizebox{0.5\textwidth}{!}{\includegraphics{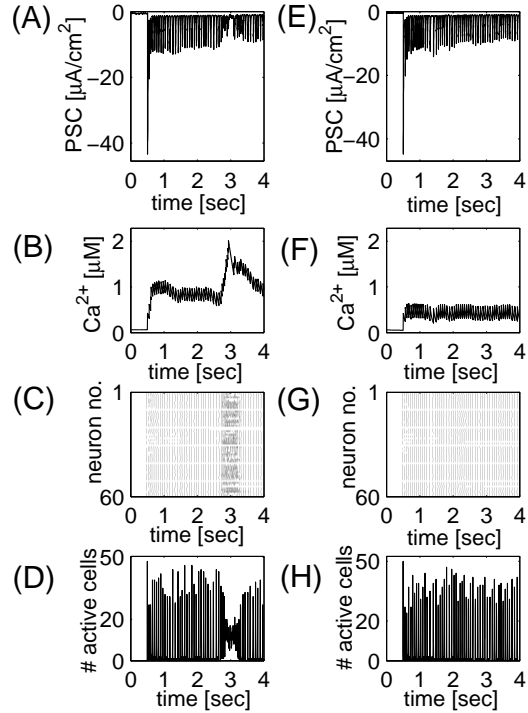}}}
\caption[]{\footnotesize The generic effect of an additional low
affinity calcium pump in regulating network reverberation. A) With
a single high affinity calcium pump (and without the low affinity
pump), reverberatory activity can be accompanied by intermittent
periods of uncoordinated activity. Such uncoordinated activity is
likely to occur when asynchronous release is stronger than
stimulated release (in other words, when $\eta_{max}$ is high and
$u$ is low). B) The period of uncoordinated activity corresponds
to an increase in residual calcium (and a subsequent uncontrolled
increase in the rate of asynchronous release). C) Network activity
during periods of high residual calcium is marked by high-rate
uncoordinated discharges of model neurons (temporal resolution is
$5 msec$). D) The coordination measure, defined as a number of
active neurons in a time window of 5 msec. Note the dip in the
number of coordinated neurons when presynaptic calcium is high. E)
The addition of a low affinity, high capacity pump eliminates
uncoordinated outbursts. To achieve this, we have modified
equation \ref{eq:eq4}, as explained in the text. This modification
keeps the level of residual calcium in a range that enables the
network to sustain reverberatory activity. F) The profile of
presynaptic calcium for the synapse simulated with an additional
pump. G) Raster plot of network activity after the low affinity
calcium pump is introduced into the model. H) The coordination
measure for the network augmented with an additional
pump.}\label{fig-fig9}
\end{figure}

Our explanation of persistent activity in small neuronal circuits
is based on an approximation of the biological complexity of the
synapse. For example, our model incorporates only a single calcium
pump, which is especially effective at clearing low $Ca^{2+}$
concentrations. Real synapses contain many more kinds of pumps and
also $Ca^{2+}$-storing organelles (such as mitochondria and ER).
The low affinity of mitochondria for $Ca^{2+}$ $(k_{m}]\sim 10\mu
M)$ is particularly suitable for regulation of $Ca^{2+}$ under
conditions when large influxes of calcium occur and other
homeostatic processes are unable to regulate $Ca^{2+}$ levels
\cite{FrielLNP}. This would be especially relevant when a network
operates in a regime of uncoordinated activity (see figure
\ref{fig-fig9}). More detailed models and experimental studies may
ultimately elucidate the role of pre-synaptic $Ca^{2+}$-storing
organelles in the regulation of persistent network activity.
Nevertheless, our results indicate that calcium regulation at the
synaptic terminal could play an important role in modulating the
dynamic activity of neuronal circuits.
\\
{\it Acknowledgements: }The authors would like to thank Eugene
Izhikevich and Nadav Raichman for valuable comments on an earlier
version of this manuscript. This work has been supported in part
by the NSF-sponsored Center for Theoretical Biological Physics
(grant numbers PHY-0216576 and PHY-0225630), by the Israeli
Science Foundation, by the Tauber Fund at Tel-Aviv University, and
by NIMH (R01 MH066962).
\newpage

\begin{appendix}
\end{appendix}

\bibliography{reverb}


\newpage

\newpage

\newpage

\newpage

\newpage

\newpage

\newpage

\newpage

\newpage

\end{document}